\documentclass[aps,prl,twocolumn,superscriptaddress,nofootinbib,10pt]{revtex4-2}

\usepackage[T1]{fontenc}
\usepackage[utf8]{inputenc}
\usepackage{lmodern}
\usepackage{amsmath,amssymb,amsfonts,bm,mathtools}
\usepackage{physics}
\usepackage{bbm}
\usepackage[colorlinks=true,allcolors=blue]{hyperref}
\usepackage{microtype}
\usepackage{graphicx}
\usepackage{placeins}
\usepackage[normalem]{ulem}

\newcommand{\ii}{\mathrm{i}}
\newcommand{\ee}{\mathrm{e}}
\newcommand{\id}{\mathbbm{1}}
\newcommand{\avg}[1]{\overline{#1}}
\newcommand{\hc}{\mathrm{H.c.}}

\begin{document}

\title{Topological protection of local quantum Fisher information}

\author{Marcin P{\l}odzie\'n}
\affiliation{Qilimanjaro Quantum Tech, Carrer de Vene\c{c}uela 74, 08019 Barcelona, Spain}

\author{Jan Chwede\'nczuk}
\affiliation{Faculty of Physics, University of Warsaw, ulica Pasteura 5, 02-093 Warszawa, Poland}

\begin{abstract}
    In many-body quantum systems, unitary dynamics generically delocalize locally encoded information, causing single-site metrological sensitivity to vanish.
    We analytically demonstrate that a topological phase can prevent this dispersal. In the open Kitaev chain, a Majorana zero mode fixes the boundary quantum Fisher information (QFI) 
    at a nonzero plateau that persists for times exponentially long in system size. We derive exact analytical expressions for the local QFI and identify the mechanism 
    as the spatial separation of the two Majorana quadratures to opposite ends of the chain. 
    This separation produces a boundary encoding-axis asymmetry that distinguishes topological boundary memory from a generic localized subgap signal. 
    We show numerically that the asymmetry is robust to moderate quenched on-site disorder, while the boundary plateau remains visible under parity-preserving interactions in finite-size real-time simulations. 
    The protocol requires only product-state initialization, Hamiltonian evolution, and single-site readout.
\end{abstract}

\maketitle

{\it Introduction.---}When classical information is encoded locally into a many-body quantum system, unitary dynamics typically 
delocalize it across the entire Hilbert space. In interacting chaotic systems, the encoded parameter becomes scrambled within highly complex many-body correlations. 
Even in exactly solvable, noninteracting systems, ballistic propagation and dephasing typically redistribute the signal into nonlocal degrees of freedom, rendering it inaccessible to any single-site probe. 
This loss of locally accessible sensitivity is quantified by the decay of local quantum Fisher information (QFI). 
Although the global evolution is unitary and preserves total information, the metrological power 
of a single-site reduced state is generally lost~\cite{Braunstein1994,Paris2009,Pezze2018,Toth2014,Zanardi2008,Hauke2016,Pezze2017,Swingle2018,Hayden2007,Plodzien2025,Calabrese2005,Jafari2020}.

Here, we demonstrate how topology can prevent complete dispersal of the local QFI. We consider a minimal encode--evolve--measure protocol in an anisotropic XY chain with open boundaries. 
A continuous parameter, $\theta$, is imprinted by a local spin rotation on a single site of a product state. 
The chain undergoes Hamiltonian evolution, and the QFI is read out from one boundary spin. Under the Jordan--Wigner mapping, the XY chain becomes the Kitaev chain, 
a paradigmatic one-dimensional topological superconductor whose topological phase hosts exponentially localized Majorana edge modes~\cite{Kitaev2001,Nayak2008,Alicea2012,Mourik2012,Hasan2010}. 
We derive a
closed-form expression for the single-site QFI in terms of Bogoliubov
propagators and show that, at the optimal operating point, it collapses to the
modulus squared of a single combined propagator. In the topological phase, the
Majorana zero mode contributes a stationary term to this propagator. Bulk
contributions dephase under time averaging, but the zero-mode term survives in
the thermodynamic limit and for exponentially long times at finite size. The
boundary QFI is therefore pinned to a nonzero long-time plateau---a
\emph{topological metrological memory} for locally encoded parameter
sensitivity.

A plateau alone, however, is not uniquely topological. Ordinary localized
boundary modes can also leave a residual long-time signal. The decisive
signature is instead a \emph{boundary encoding-axis asymmetry}, rooted in
the spatial separation of the two Majorana modes to opposite ends of the
chain. Under the Jordan--Wigner mapping, different local spin-rotation axes
at a boundary site project onto different linear combinations of the
Bogoliubov amplitudes $u_0$ and $v_0^*$ of the zero mode. One combination,
$u_0+v_0^*$, defines the Majorana mode localized at that boundary and carries
$O(1)$ weight; the orthogonal combination, $u_0-v_0^*$, defines the Majorana
mode at the opposite end and is exponentially suppressed in $L/\xi$, where $L$ is the chain size and $\xi$ is the localization length.
For the real-pairing convention used below and positive anisotropy of the XY chain ($\gamma>0$), encoding information
via a local rotation through the $y$-axis at
the left boundary couples to the local Majorana and produces the full QFI
plateau. On the other hand, $x$-axis encoding projects onto the distant mode and is
exponentially suppressed. A trivial localized boundary mode can also produce a residual plateau, and may even show accidental quadrature polarization. What distinguishes the topological case is the parametrically large and boundary-correlated asymmetry: one quadrature has $O(1)$ weight at the left edge, while the orthogonal quadrature is suppressed by the overlap with the opposite-edge Majorana and therefore decays exponentially with system size.
We confirm the boundary-axis interpretation in the presence of quenched on-site disorder
and show that the associated boundary plateau remains visible under symmetry-preserving nearest-neighbor interactions that
break the quadratic free-fermion structure and, for generic parameters, destroy integrability.
Numerically, the asymmetry remains close to its clean value below the clean bulk-gap scale and degrades at stronger disorder, where the residual boundary signal becomes consistent with ordinary localized subgap structure.

\emph{Model and protocol.}---We consider the spin-$\frac{1}{2}$ anisotropic XY
chain
\begin{align}
  \hat H= -\frac{J}{2}\sum_j \big[(1{+}\gamma)\hat\sigma^x_j\hat\sigma^x_{j+1}
    + (1{-}\gamma)\hat\sigma^y_j\hat\sigma^y_{j+1}\big]- h\sum_j \hat\sigma^z_j,
  \label{eq:HXY}
\end{align}
characterized by the hopping constant $J$, the anisotropy $\gamma$ and the transverse field $h$. 
This Hamiltonian maps via the Jordan--Wigner transformation
\begin{align}
  \hat c_j=(\prod_{m<j}\hat\sigma_m^z)\hat\sigma_j^-
\end{align}
to the Kitaev chain~\cite{Kitaev2001,Lieb1961},
\begin{align}
  \hat H_{\rm K}= \sum_j\left[J(\hat c_j^\dagger \hat c_{j+1}{+}\hc)-\gamma J(\hat c_j \hat c_{j+1}{+}\hc)- 2h\hat n_j\right],
  \label{eq:Hferm}
\end{align}
where $\hat n_j=\hat c^\dagger_j\hat c_j$. For real couplings the model belongs to symmetry class
BDI~\cite{Ryu2010,Chiu2016}. The bulk dispersion is
\begin{equation}\label{eq:dispersion}
  \varepsilon_q = 2\sqrt{(h{-}J\cos q)^2+J^2\gamma^2\sin^2 q},
\end{equation}
and the bulk gap closes at $|h|=J$ for $\gamma\neq0$, separating the
topological phase $(|h|<J)$ from the trivial one $(|h|>J)$.

Because Eq.~\eqref{eq:Hferm} is quadratic, the Heisenberg-evolved fermion
operator takes the Bogoliubov form
\begin{equation}
\hat c_j(t)=\sum_\ell\!\left[U_{j,\ell}(t)\hat c_\ell+V_{j,\ell}(t)\hat c_\ell^\dagger\right],
\label{eq:cjt}
\end{equation}
where, in the Bogoliubov--de Gennes (BdG) eigenbasis with mode functions $(u_\nu(j),v_\nu(j))$
(chosen real for the BDI chain),
\begin{align}
U_{j,\ell}(t) &= \sum_\nu\!\left[u_\nu(j)u_\nu^*(\ell)\,\ee^{-\ii\varepsilon_\nu t}
+ v_\nu^*(j)v_\nu(\ell)\,\ee^{+\ii\varepsilon_\nu t}\right], \label{eq:Ujl}\\
V_{j,\ell}(t) &= \sum_\nu\!\left[u_\nu(j)v_\nu^*(\ell)\,\ee^{-\ii\varepsilon_\nu t}
+ v_\nu^*(j)u_\nu(\ell)\,\ee^{+\ii\varepsilon_\nu t}\right]. \label{eq:Vjl}
\end{align}
All dynamical information entering the single-site QFI is encoded in these two
propagators.

We initialize the system in the fully polarized product state
$|\!\downarrow\cdots\downarrow\rangle$, i.e. the Jordan--Wigner vacuum
$|0\rangle$, and imprint the unknown angle $\theta$ at site $k$ through the
local rotation. We consider two ``channels'', corresponding to two orthogonal rotation axes, namely
\begin{align}
  \hat R_{x/y}^{(k)}(\theta)=e^{-\ii\theta\hat\sigma_k^{x/y}/2}.
\end{align} 
For the $y$ case, and in the fermionic convention used below, we write
\begin{equation}
|\Psi(\theta)\rangle
=
\cos\tfrac{\theta}{2}\,|0\rangle
-
\sin\tfrac{\theta}{2}\,\hat c_k^\dagger|0\rangle.
\label{eq:init}
\end{equation}
In the derivation below we use the fermionic convention
$|k\rangle\equiv \hat c_k^\dagger|0\rangle$. For a physical spin rotation
at $k>1$, the Jordan--Wigner string contributes an additional factor
$(-1)^{k-1}$ to the one-particle component. This only multiplies the
transverse Bloch vector, and hence $W_{j,k}$, by an overall sign and drops
out of the QFI.
The system then evolves under $\hat H$ from Eq.~\eqref{eq:HXY}. We read out the QFI from the left boundary
site, $j=1$, where the Jordan--Wigner string is absent and the measurement is
strictly local in the spin language. Away from the boundary the fermionic
single-site algebra maps to string-dressed spin operators, so the operationally
local protocol is the boundary one studied here.

\emph{Single-site QFI.}---The reduced state of site $j$ is a qubit with a density matrix
$\hat\varrho_j=\frac12(\id+\bm m\cdot\bm\hat\sigma)$ and the information about $\theta$, stored in $\bm m$, is quantified by the QFI
\begin{equation}\label{eq:QFI_qubit}
  F_Q = |\partial_\theta\bm m|^2 +\frac{(\bm m\cdot\partial_\theta\bm m)^2}{1-|\bm m|^2}.
\end{equation}
This formula stems from the standard expression for the QFI, see~\cite{Braunstein1994,supmat}. Because the fermionic Hamiltonian is
a quadratic form of the operators $\hat c_i$, the Bogoliubov propagators give an exact expression for the Bloch vector $\bm m$, i.e., 
\begin{align}\label{eq:mperp}
  m_x+i m_y= -\sin\theta\,W_{j,k}(t), \ \ \ m_z= Z_0+\Delta p_{j,k}\cos\theta, 
\end{align}
where the overall sign in the transverse component is immaterial for the QFI,
with
\begin{subequations}  \label{eq:Zj}
  \begin{align}
    &W_{j,k}\equiv U_{j,k}+V_{j,k},\qquad\Delta p_{j,k}=|U_{j,k}|^2-|V_{j,k}|^2,\\
    &S_j=\sum_\ell |V_{j,\ell}|^2,\qquad Z_0=1-2S_j-\Delta p_{j,k}.
  \end{align}
\end{subequations}
Substituting Eqs.~\eqref{eq:mperp}--\eqref{eq:Zj} into
Eq.~\eqref{eq:QFI_qubit} gives the closed-form single-site QFI
\begin{align}
F_Q^{(j)} &= |W|^2\cos^2\theta+\Delta p^2\sin^2\theta \nonumber\\
&\quad +\frac{\sin^2\theta\,[\cos\theta(|W|^2{-}\Delta p^2)-\Delta p\,Z_0]^2}
{1-|W|^2\sin^2\theta-(Z_0+\Delta p\cos\theta)^2}.
\label{eq:QFI_full}
\end{align}
This exact expression shows that the local QFI is fully determined by two-point propagators.

The structure simplifies sharply at the optimal operating point $\theta_0=0$.
There the Bloch vector is purely longitudinal, while its parameter derivative is
purely transverse, so the mixed-state correction in Eq.~\eqref{eq:QFI_qubit}
vanishes identically, giving
\begin{equation}
F_Q^{(j)}\big|_{\theta_0=0}=|W_{j,k}(t)|^2.
\label{eq:QFI_simple}
\end{equation}
This geometric simplification reflects parity-sector protection: at
$\theta_0{=}0$ the encoded parameter resides entirely in the parity-odd
coherence channel, which is decoupled from the parity-even pairing
fluctuations by fermion-parity conservation.
Since the QFI is the maximum classical Fisher information over all local measurements on the reduced qubit, $|W_{j,k}|^2$ is the locally extractable small-signal sensitivity at $\theta_0=0$. More generally,
$F_Q^{(j)}\ge\cos^2\theta_0\,|W_{j,k}(t)|^2$, so the plateau discussed
below persists away from the optimal operating point.

\begin{figure}[!t]
\centering
\includegraphics[width=0.95\columnwidth]{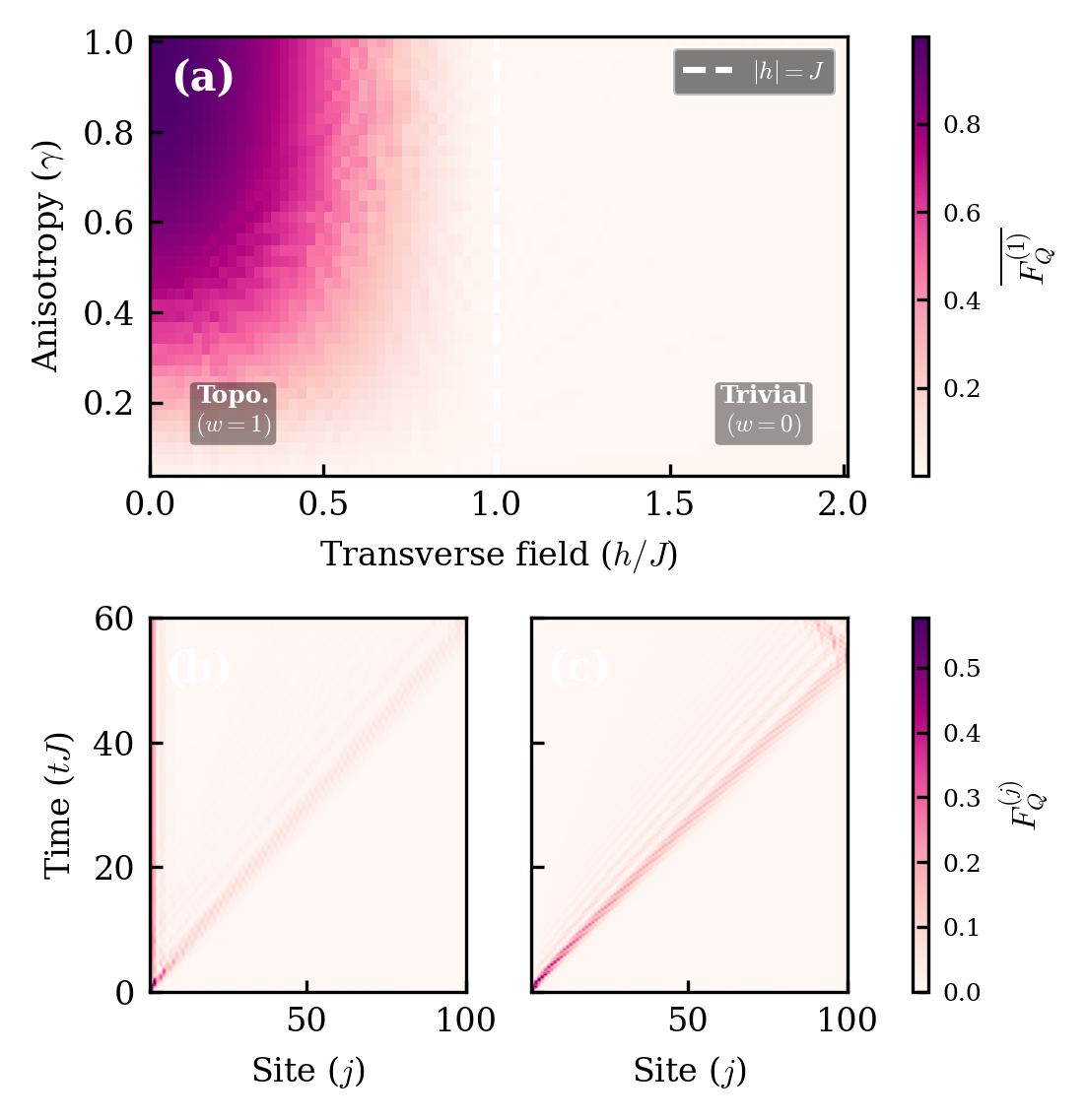}
\caption{%
(a)~Time-averaged boundary QFI $\avg{F_Q^{(1)}}$ across the $(h/J,\gamma\ge0)$ phase diagram
(boundary encoding and readout, $k{=}j{=}1$,
averaging window $tJ\in[150,200]$); dashed line: $|h|{=}J$.
(b)~Topological phase ($\gamma{=}0.3$, $h{=}0.5J$): ballistic spreading
coexists with a persistent boundary streak from the Majorana zero mode.
(c)~Trivial phase ($\gamma{=}0.8$, $h{=}1.8J$): complete dephasing with no
persistent boundary signal.
Here: $L{=}100$, $k{=}1$, $\theta_0{=}0$.}
\label{fig:hero}
\end{figure}

\emph{Topological plateau.}---In the topological phase, the open chain supports
a near-zero-energy mode. Choosing the zero-mode BdG eigenvector
$(u_0(j),v_0(j))$ real, the two Majorana envelopes are
\begin{equation}
\phi_L(j)=\frac{u_0(j)+v_0(j)}{\sqrt2},\qquad
\phi_R(j)=\frac{u_0(j)-v_0(j)}{\sqrt2},
\label{eq:majorana_LR}
\end{equation}
localized at opposite boundaries and satisfying the normalization $\sum_j \phi_L(j)^2=\sum_j\phi_R(j)^2=\frac12$. For the real-coupling chain and $\gamma>0$,
the two orthogonal local encoding channels are
\begin{equation}
W^{(y)}=U+V,\qquad W^{(x)}=U-V.
\label{eq:WRyRx}
\end{equation}
Their zero-mode contributions are
\begin{align}\label{eq:W_zm_Ry}
  W_{j,k}^{(y/x)}(t)\big|_0=2\phi_{L/R}(k)\big[\phi_{L/R}(j)\alpha-i\,\phi_{R/L}(j)\beta\big],
\end{align}
where $\alpha=\cos(\varepsilon_0 t)$ and $\beta=\sin(\varepsilon_0 t)$.
These expressions make the boundary-axis asymmetry explicit. At the left
boundary, the $\hat R_y$ channel couples to the left Majorana with $O(1)$ weight,
whereas the $\hat R_x$ channel is controlled by the exponentially small left-boundary
weight of the opposite-edge Majorana. In the thermodynamic or semi-infinite
limit, this asymmetry becomes exact; at large finite $L$, the orthogonal
boundary response is exponentially suppressed in $L/\xi$.

Under long-time averaging within the window
$\varepsilon_{\rm gap}^{-1}\ll T_{\rm avg}\ll\varepsilon_0^{-1}$, the
oscillatory bulk contributions dephase for generic open-chain spectra, while the
zero-mode contribution remains quasi-stationary~\cite{supmat}. Using Eq.~\eqref{eq:QFI_simple}, the
time-averaged boundary QFI therefore takes the form
\begin{equation}
\avg{F_Q^{(j)}}\big|_{\theta_0=0}
=
\avg{|W^{\rm bulk}_{j,k}|^2}
+
4|\phi_L(j)\phi_L(k)|^2.
\label{eq:plateau_exact}
\end{equation}
The second term is a nonzero edge plateau pinned by the Majorana zero mode. For
boundary encoding and boundary readout, $j=k=1$, the normalization of the
exponential Majorana envelope implies
$4|\phi_L(1)|^4\sim\xi^{-2}\sim |h-J|^2$ as the transition is approached at
fixed $\gamma\neq0$. The plateau therefore vanishes continuously at criticality,
while remaining nonzero throughout the topological phase.

In the trivial phase, no zero mode exists. Extended bulk modes carry boundary
amplitude $\sim 1/\sqrt{L}$, so the long-time boundary QFI vanishes as
$O(1/L)$ in the thermodynamic limit. The plateau thus requires both a pairing
gap and a topological edge mode. In the gapless XX limit ($\gamma=0$), the
anomalous propagator vanishes identically and no plateau forms.

\emph{Results.}---Figure~\ref{fig:hero}(a) maps the time-averaged boundary QFI
$\avg{F_Q^{(1)}}$ across the $(h/J,\gamma)$ plane for $\gamma\ge0$, $L=100$, and
$\theta_0=0$. A robust nonzero plateau fills the topological sector
$(|h|<J,\gamma\neq0)$ and is strongly suppressed in the trivial sector, with the
crossover aligned with the bulk phase boundary at $|h|=J$. The corresponding
real-time dynamics are shown in Fig.~\ref{fig:hero}(b,c). In the topological
phase, ballistic spreading at the maximum BdG group velocity coexists with a persistent boundary streak, whereas in
the trivial phase the boundary signal dephases completely~\cite{supmat}.

\begin{figure}[!t]
\centering
\includegraphics[width=0.95\columnwidth]{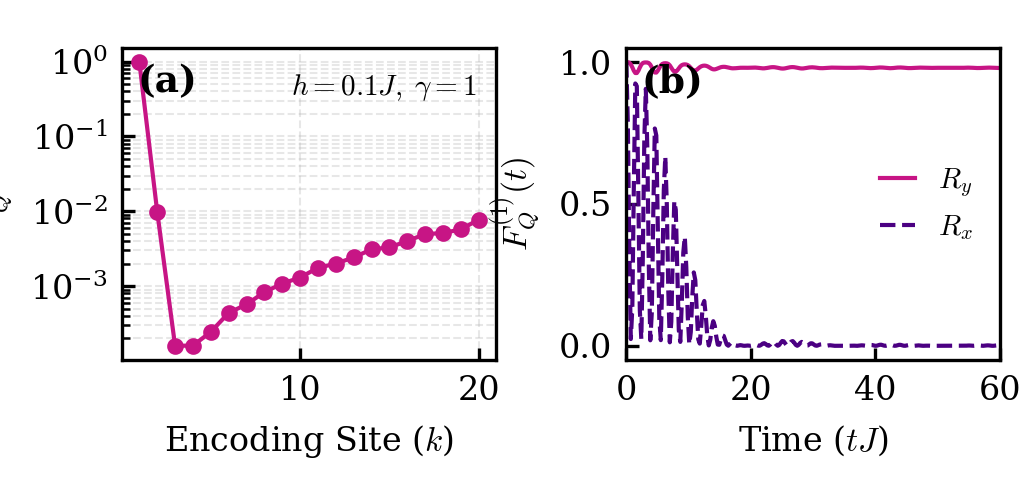}
\caption{%
Mechanism of the boundary metrological plateau ($L{=}100$, $\gamma{=}1$,
$h{=}0.1J$, $j{=}1$, averaging window $tJ\in[150,200]$).
(a)~Encoding-site scan: the time-averaged boundary QFI decays as
$\propto e^{-2(k-1)/\xi}$, directly resolving the Majorana localization length.
(b)~Boundary encoding-axis asymmetry: for left-boundary readout and $\gamma>0$,
the $\hat R_y$ channel sustains the plateau, while the orthogonal $\hat R_x$ channel is
strongly suppressed and vanishes in the thermodynamic limit.}
\label{fig:mechanics}
\end{figure}

\emph{Topological mechanism.}---The plateau is controlled by the spatial
structure of the zero mode. Moving the encoding site into the bulk suppresses
the time-averaged boundary QFI as $\exp[-2(k-1)/\xi]$, directly tracking the
left-edge Majorana envelope [Fig.~\ref{fig:mechanics}(a)].

The boundary-axis asymmetry should be understood as an asymmetry of the coherence prepared by the encoding rotation. With the convention $\hat c_j=(\prod_{m<j}\hat\sigma_m^z)\hat\sigma_j^-$, the boundary Majorana quadratures are $\hat X_1=\hat c_1+\hat c_1^\dagger$ and $\hat Y_1=-i(\hat c_1-\hat c_1^\dagger)$. A rotation of the initial $|\!\downarrow\cdots\downarrow\rangle$ state about the physical $y$-axis prepares real coherence and therefore probes $\hat X_1$, giving the combined propagator $W^{(y)} = U+V$. A rotation about the physical $x$-axis prepares imaginary coherence and probes the orthogonal quadrature, giving $W^{(x)} = U-V$, up to an irrelevant phase. For $\gamma>0$, $U+V$ overlaps with the left Majorana mode, while $U-V$ overlaps with the right Majorana mode whose weight at the left edge is exponentially small [Fig.~\ref{fig:mechanics}(b)]. The asymmetry is a direct consequence of the topological separation of the two Majorana modes to opposite edges, not of boundary localization alone.

\begin{figure}[!t]
\centering
\includegraphics[width=0.8\columnwidth]{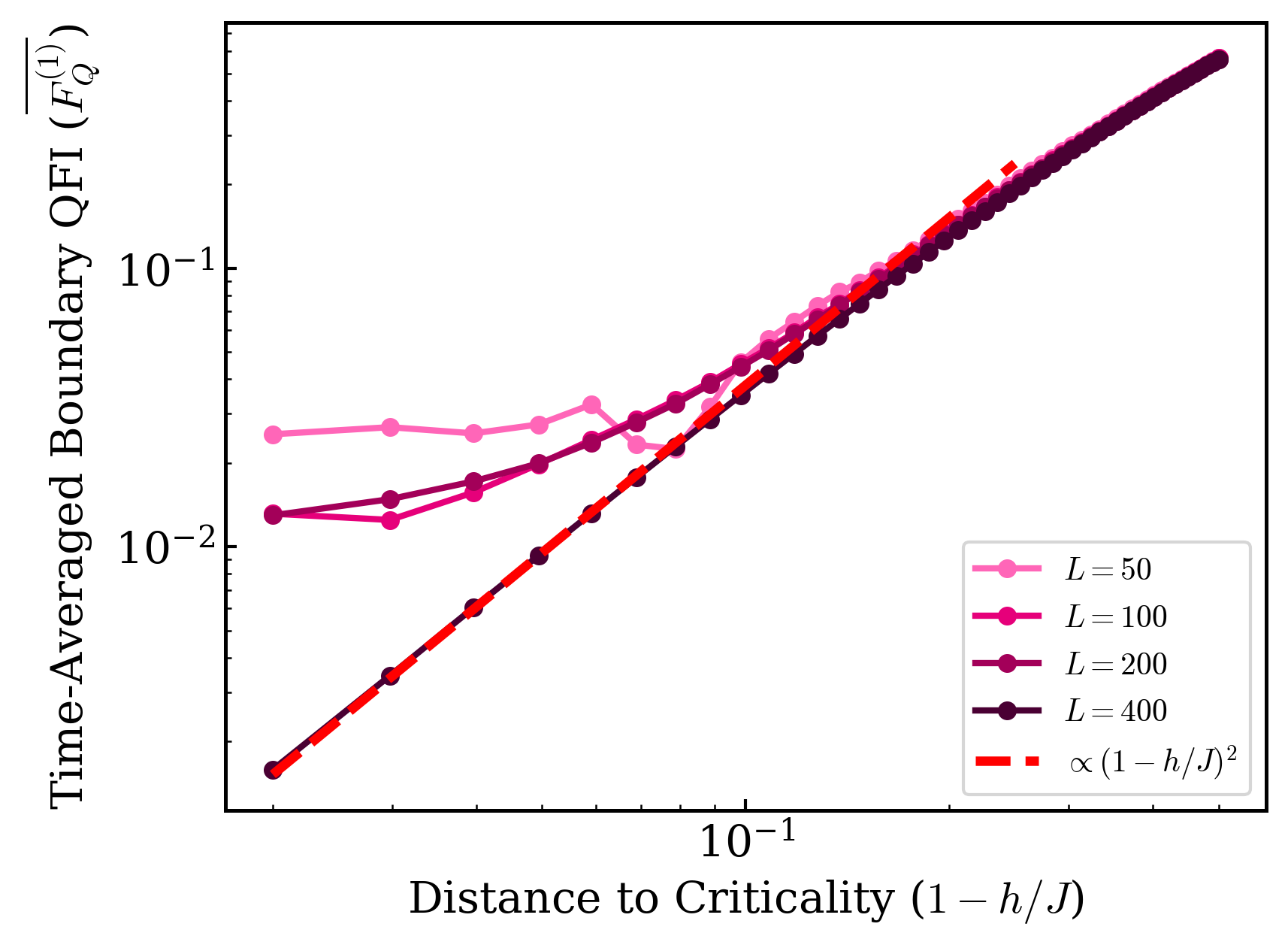}
\caption{%
Critical scaling of the boundary QFI near the topological transition. For fixed
$\gamma{=}1$ and boundary encoding/readout, the data for
$L\in\{50,100,200,400\}$ approach the asymptotic scaling
$\avg{F_Q^{(1)}}\propto(1-h/J)^2$ (dashed line). Parameters: $\theta_0{=}0$.}
\label{fig:scaling}
\end{figure}

\emph{Critical scaling.}---The plateau amplitude behaves as a dynamical order
parameter for the topological transition. For boundary encoding and boundary
readout, Eq.~\eqref{eq:plateau_exact} gives
$\avg{F_Q^{(1)}}\propto|\phi_L(1)|^4$. At fixed $\gamma\neq0$, the Majorana
normalization implies $|\phi_L(1)|^2\sim\xi^{-1}$, while
$\xi\sim|h-J|^{-1}$ on approach to the transition. Hence
\begin{equation}
\avg{F_Q^{(1)}} \sim |h-J|^2,
\label{eq:critical_scaling}
\end{equation}
for $h\to J$ at fixed nonzero anisotropy. Figure~\ref{fig:scaling} confirms
this quadratic scaling for increasing system sizes, demonstrating convergence
toward the thermodynamic-limit prediction. Very close to the transition, finite-size and finite-time effects round the scaling because the bulk dephasing time diverges as the gap closes.

\begin{figure}[!t]
\centering
\includegraphics[width=\columnwidth]{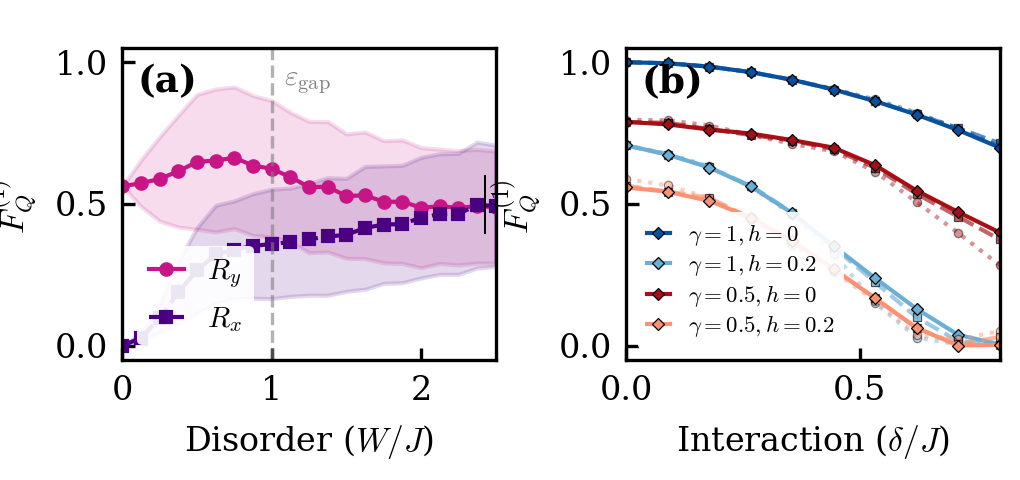}
\caption{%
Robustness of the boundary metrological plateau.
(a)~Disorder: disorder-averaged boundary QFI versus disorder strength $W/J$
($h{=}0.5J$, $\gamma{=}1$, $L{=}200$, $N_{\rm dis}{=}1000$). The $\hat R_y$
channel (pink) preserves the plateau for $W\lesssim\varepsilon_{\rm gap}$
while $\hat R_x$ (purple) remains suppressed; shaded bands show one standard
deviation. For disorder strengths above the clean bulk-gap scale, both channels converge.
(b)~Interactions: time-averaged boundary QFI versus interaction strength
$\delta/J$ for $\gamma{=}1$ (blue), $\gamma{=}0.5$ (red), at $h{=}0$ (dark)
and $h{=}0.2J$ (light). System sizes $L{=}12$ (dotted), $16$ (dashed), $20$
(solid) show weak size dependence. Parameters: $tJ\in[25,50]$ averaging window,
$\hat R_y$ encoding, $k{=}j{=}1$.}
\label{fig:robustness}
\end{figure}

\emph{Disorder robustness.}---A localized boundary signal alone does not
distinguish topology from ordinary localization, but the boundary-axis
asymmetry does. We therefore test its robustness against static on-site
disorder, $h_j=h+\delta h_j$ with $\delta h_j\in[-W,W]$. Figure
\ref{fig:robustness}(a) shows the disorder-averaged long-time boundary QFI for the
two orthogonal encoding channels. For disorder strengths below the clean bulk-gap scale, the data show that the boundary-axis asymmetry remains close to its clean value. This is consistent with the Majorana interpretation: moderate disorder deforms the zero-mode envelope without bringing both Majorana components to the same boundary. At stronger disorder, the two encoding channels become comparable, indicating that the residual boundary response is no longer uniquely tied to the clean topological Majorana mechanism.

\emph{Interacting regime.}---The preceding analysis is exact for the
quadratic chain. In the interacting case, the exact free-fermion identity $F_Q=|W|^2$ no longer applies. Topological ground-state edge degeneracy is expected to remain stable under parity-preserving perturbations that do not close the bulk gap, but the persistence of a dynamical boundary-QFI plateau after a product-state quench is a stronger finite-energy statement. We therefore test it numerically by adding a nearest-neighbor Ising interaction $\delta\sum_j\hat\sigma^z_j\hat\sigma^z_{j+1}$ to the XY Hamiltonian. This coupling breaks the quadratic free-fermion structure and, for generic parameters, destroys integrability. It couples the edge sector to many-body scattering continua, yet it respects the $\mathbb{Z}_2$ symmetry.

Exact statevector simulations on chains consisting of $L$ from $12$ to $20$ spins
[Fig.~\ref{fig:robustness}(b)] show that the boundary plateau remains visible
deep into the interacting regime over the accessible sizes and time windows. At the Kitaev sweet spot ($h{=}0$,
$\gamma{=}1$), where the topological gap is maximal, the boundary QFI
retains ${\sim}70\%$ of its free-fermion value at $\delta{=}0.8J$. The
interaction dresses the edge mode with many-body correlations and
renormalizes the bulk spectrum. As long as the gap remains finite, the numerical data are consistent with an adiabatically dressed edge mode that retains substantial boundary weight over the accessible time window. For finite transverse field
($h{=}0.2J$), the bare gap is smaller and the system sits closer to the
quantum critical point, so interactions erode the plateau more rapidly by
facilitating gap narrowing and enhanced boundary scattering. The weak size dependence across
$L=12, 16, 20$ supports the interpretation that the interacting plateau is not a small-system artifact.

\emph{Discussion.}---In the topological phase, the Majorana
zero mode pins the boundary QFI to a nonzero long-time plateau, and the
boundary-axis asymmetry identifies its topological origin.
The plateau forms on the bulk-dephasing timescale $\sim\varepsilon_{\rm
gap}^{-1}$ and persists for exponentially long times $\sim e^{L/\xi}$ before
finite-size Majorana hybridization becomes relevant. Because the protocol
starts from a fully polarized product state, it is naturally implemented as a
quench experiment requiring no ground-state preparation. The required
ingredients---anisotropic spin exchange, local rotations, quench dynamics, and single-site readout---are available or can be engineered in Rydberg atom
arrays~\cite{Bernien2017}, superconducting circuits~\cite{Roushan2017}, and
trapped-ion chains~\cite{Garttner2017}, for example through Floquet
engineering or digital-analog decomposition.
In such implementations, distinguishing a genuine topological phase from a
trivially localized boundary mode is a practical challenge; for example,
zero-bias conductance peaks can be mimicked by trivial Andreev bound
states~\cite{Mourik2012}. The encoding-axis
asymmetry provides a complementary local diagnostic tool that resolves this
ambiguity with strictly local measurements, without
requiring nonlocal correlators or parameter sweeps across the phase boundary.
As shown in Fig.~\ref{fig:robustness}(b), the plateau remains visible under
symmetry-preserving interactions over the accessible sizes and time windows, consistent with the stability of the
Majorana edge mode~\cite{Fidkowski2011,Zeng2019}, although the exact free-fermion identity
$F_Q{=}|W|^2$ at $\theta_0=0$ no longer applies. The combination of a topologically
controlled plateau, a boundary-axis diagnostic, and finite-size evidence for the persistence of the plateau under interactions constitutes an operational metrological manifestation of
bulk--boundary correspondence.

\begin{acknowledgments}
M.P. acknowledges RES resources provided by Barcelona Supercomputing Center in Marenostrum 5 to NNO-2025-3-0004.
\end{acknowledgments}

\bibliographystyle{apsrev4-2}
\bibliography{references}


\FloatBarrier
\clearpage
\onecolumngrid
\appendix
\setcounter{equation}{0}
\setcounter{figure}{0}
\renewcommand{\theequation}{S\arabic{equation}}
\renewcommand{\thefigure}{S\arabic{figure}}
\renewcommand{\thesection}{S\Roman{section}}

\begin{center}
\textbf{\large Supplemental Material: Topological protection of local quantum
Fisher information}
\end{center}

\section{Jordan--Wigner mapping and BdG framework}
\label{sm:model}

The spin-$\frac12$ anisotropic XY chain
\begin{equation}
\hat H = -\frac{J}{2}\sum_j \big[(1{+}\gamma)\hat\sigma^x_j\hat\sigma^x_{j+1}
+ (1{-}\gamma)\hat\sigma^y_j\hat\sigma^y_{j+1}\big] - h\sum_j \hat\sigma^z_j
\label{eq:sm_HXY}
\end{equation}
maps under the Jordan--Wigner transformation
\begin{equation}
\hat c_j=\Big(\prod_{m<j}\hat\sigma_m^z\Big)\hat\sigma_j^-,
\qquad
\hat c_j^\dagger=\hat\sigma_j^+\Big(\prod_{m<j}\hat\sigma_m^z\Big),
\label{eq:sm_JW}
\end{equation}
to the Kitaev chain
\begin{equation}
\hat H = J\sum_j (\hat c_j^\dagger \hat c_{j+1}+\hc)
- J\gamma\sum_j (\hat c_j \hat c_{j+1}+\hc)
- 2h\sum_j \hat n_j.
\label{eq:sm_Hferm}
\end{equation}
Here $J$ is the hopping amplitude, $\Delta=J\gamma$ is the $p$-wave pairing
strength, and $\mu=-2h$ is the chemical potential. For real couplings, the
model belongs to symmetry class BDI~\cite{Ryu2010,Chiu2016}. The bulk gap closes
at $|h|=J$ for $\gamma\neq0$, separating the topological phase
$(|h|<J)$ from the trivial phase $(|h|>J)$.

On the infinite chain, introducing the Nambu spinor
$\hat\Psi_q=(\hat c_q,\hat c_{-q}^\dagger)^{\mathsf T}$ gives the Bogoliubov--de~Gennes (BdG)
form
\begin{equation}
  \hat H=\frac12\int_{-\pi}^{\pi}\frac{dq}{2\pi}\,\hat\Psi_q^\dagger \hat H_q\hat\Psi_q+\mathrm{const.},
\label{eq:sm_HBdG}
\end{equation}
with
\begin{equation}
  \hat H_q=a_q\hat\tau_z+b_q\hat\tau_y,
\qquad
a_q=-2(h-J\cos q),
\qquad
b_q=2J\gamma\sin q,
\label{eq:sm_Hq}
\end{equation}
where $\hat\tau_{y,z}$ are Pauli matrices acting in Nambu (particle--hole) space,
and the
quasiparticle dispersion is
\begin{equation}
\varepsilon_q=\sqrt{a_q^2+b_q^2}
=
2\sqrt{(h-J\cos q)^2+J^2\gamma^2\sin^2 q}.
\label{eq:sm_dispersion}
\end{equation}
Since $\hat H_q^2=\varepsilon_q^2\,\id$, the time-evolution operator is
\begin{equation}
  \hat U_q(t)=e^{-i\hat H_q t}=\cos(\varepsilon_q t)\,\id-i\,\frac{\sin(\varepsilon_q t)}{\varepsilon_q}\,(a_q\tau_z+b_q\tau_y).
\label{eq:sm_Uq}
\end{equation}
Its matrix elements define the Bogoliubov amplitudes
\begin{align}
  u_q(t)=\cos(\varepsilon_q t)-i\frac{a_q}{\varepsilon_q}\sin(\varepsilon_q t),\ 
  v_q(t)= -\frac{b_q}{\varepsilon_q}\sin(\varepsilon_q t),
\end{align}
which satisfy $|u_q|^2+|v_q|^2=1$.

The real-space propagators are obtained by Fourier transformation,
\begin{equation}
U_r(t)=\int_{-\pi}^{\pi}\frac{dq}{2\pi}\,e^{iqr}u_q(t),
\qquad
V_r(t)=\int_{-\pi}^{\pi}\frac{dq}{2\pi}\,e^{iqr}v_q(t),
\label{eq:sm_UV}
\end{equation}
so that
\begin{equation}
\hat c_j(t)=\sum_\ell\big[U_{j-\ell}(t)\hat c_\ell+V_{j-\ell}(t)\hat c_\ell^\dagger\big].
\label{eq:sm_cjt_inf}
\end{equation}

For a finite open chain, the BdG Hamiltonian is diagonalized by discrete
quasiparticle modes $\gamma_\nu$ with energies $\varepsilon_\nu\ge0$ and mode
functions $(u_\nu(j),v_\nu(j))$,
\begin{equation}
\hat c_j=\sum_\nu\left[u_\nu(j)\gamma_\nu+v_\nu^*(j)\gamma_\nu^\dagger\right].
\label{eq:sm_modeexp}
\end{equation}
The corresponding spectral propagators are
\begin{align}
U^{\rm OBC}_{j,k}(t)
&=
\sum_\nu\big[u_\nu(j)u_\nu^*(k)e^{-i\varepsilon_\nu t}
+v_\nu^*(j)v_\nu(k)e^{+i\varepsilon_\nu t}\big],
\label{eq:sm_UOBC}\\
V^{\rm OBC}_{j,k}(t)
&=
\sum_\nu\big[u_\nu(j)v_\nu^*(k)e^{-i\varepsilon_\nu t}
+v_\nu^*(j)u_\nu(k)e^{+i\varepsilon_\nu t}\big].
\label{eq:sm_VOBC}
\end{align}
For the real-coupling BDI chain considered in the main text, the mode functions
may be chosen real.

\section{Single-site QFI}
\label{sm:qfi}

\subsection{QFI of a single qubit}

For a density matrix $\hat\varrho(\theta)$ with spectral decomposition
$\hat\varrho=\sum_n\lambda_n|\lambda_n\rangle\langle\lambda_n|$, the quantum Fisher
information is~\cite{Braunstein1994}
\begin{equation}
F_Q=\sum_{n,m}
\frac{2\,|\langle\lambda_n|\partial_\theta\hat\varrho|\lambda_m\rangle|^2}
{\lambda_n+\lambda_m},
\label{eq:sm_QFI_eigen}
\end{equation}
where the sum is restricted to $\lambda_n+\lambda_m>0$.

For a qubit
\begin{equation}
\hat\varrho=\frac12(\id+\bm m\cdot\bm\hat\sigma),
\label{eq:sm_rho_bloch}
\end{equation}
the eigenvalues are $\lambda_\pm=(1\pm|\bm m|)/2$, with eigenvectors aligned
with $\hat{\bm m}=\bm m/|\bm m|$. Direct substitution into
Eq.~\eqref{eq:sm_QFI_eigen} gives
\begin{equation}
F_Q
=
|\partial_\theta\bm m|^2
+
\frac{(\bm m\cdot\partial_\theta\bm m)^2}{1-|\bm m|^2}.
\label{eq:sm_QFI_qubit}
\end{equation}
This is the formula used in the main text.

\subsection{Encoded state and parity structure}

The encoding step is a local spin rotation on site $k$,
\begin{equation}
\hat R_y^{(k)}(\theta)=e^{-i\theta\hat\sigma_k^y/2},
\label{eq:sm_Ry}
\end{equation}
applied to the fully polarized product state
$|\!\downarrow\cdots\downarrow\rangle$, i.e. the Jordan--Wigner vacuum
$|0\rangle$. It prepares
\begin{equation}
|\Psi(\theta)\rangle
=
\cos\frac{\theta}{2}|0\rangle
-
\sin\frac{\theta}{2}\hat c_k^\dagger|0\rangle.
\label{eq:sm_init}
\end{equation}
In the derivation below we use the fermionic convention
$|k\rangle\equiv \hat c_k^\dagger|0\rangle$. For a physical spin rotation
at $k>1$, the Jordan--Wigner string contributes an additional factor
$(-1)^{k-1}$ to the one-particle component. This only multiplies the
transverse Bloch vector, and hence $W_{j,k}$, by an overall sign and drops
out of the QFI.

For any Hermitian operator $\hat A$,
\begin{equation}
  \langle \hat A\rangle=\cos^2\frac{\theta}{2}\,\langle0|\hat A|0\rangle+\sin^2\frac{\theta}{2}\,\langle k|\hat A|k\rangle
  -\sin\theta\,\Re\langle0|\hat A|k\rangle.
\label{eq:sm_exp_decomp}
\end{equation}
Because the Hamiltonian preserves fermion parity, the time-evolved states
$e^{-i\hat Ht}|0\rangle$ and $e^{-i\hat Ht}|k\rangle$ remain in the even and odd parity
sectors, respectively. Parity-odd operators contribute only through the
interference term, while parity-even operators contribute only through the
diagonal terms.

We define the local fermionic qubit operators\footnote{At the left boundary, where the Jordan--Wigner string is absent, $\hat c_1 = \frac{1}{2}(\hat\sigma_1^x - i\hat\sigma_1^y)$. Hence $\hat Y_1 = -\hat\sigma_1^y$ and $\hat Z_1 = -\hat\sigma_1^z$. This sign convention merely reflects/rotates the Bloch vector and leaves the QFI invariant.}
\begin{equation}
  \hat X_j=\hat c_j+\hat c_j^\dagger,\qquad
  \hat Y_j=-i(\hat c_j-\hat c_j^\dagger),\qquad
  \hat Z_j=1-2\hat n_j.
\label{eq:sm_XYZ}
\end{equation}
The reduced single-site state is
$\hat\varrho_j=\frac12(\id+\bm m\cdot\bm\hat\sigma)$ with Bloch components
\begin{equation}
  m_x=\langle \hat X_j\rangle,\qquad
  m_y=\langle \hat Y_j\rangle,\qquad
  m_z=\langle \hat Z_j\rangle.
  \label{eq:sm_bloch_components}
\end{equation}

\subsection{Transverse Bloch vector and combined propagator}

The key identity is
\begin{equation}
  m_x+i m_y = 2\langle \hat c_j(t)\rangle.
  \label{eq:sm_mxy_c}
\end{equation}
Using Eq.~\eqref{eq:sm_init},
\begin{align}\label{eq:sm_c_expectation}
  \langle \hat c_j(t)\rangle  = -\frac{\sin\theta}{2}\Big(\langle0|\hat c_j(t)|k\rangle+\langle k|\hat c_j(t)|0\rangle\Big).
\end{align}
The first matrix element follows from the Heisenberg expansion:
\begin{equation}
\langle0|\hat c_j(t)|k\rangle
=
\sum_\ell U_{j,\ell}\,\langle0|\hat c_\ell|k\rangle
+
\sum_\ell V_{j,\ell}\,\langle0|\hat c_\ell^\dagger|k\rangle
=
U_{j,k},
\label{eq:sm_cross_U}
\end{equation}
since $\langle0|\hat c_\ell|k\rangle=\delta_{\ell k}$ and
$\langle0|\hat c_\ell^\dagger|k\rangle=0$.
Similarly,
\begin{equation}
\langle k|\hat c_j(t)|0\rangle
=
\sum_\ell U_{j,\ell}\,\langle k|\hat c_\ell|0\rangle
+
\sum_\ell V_{j,\ell}\,\langle k|\hat c_\ell^\dagger|0\rangle
=
V_{j,k},
\label{eq:sm_cross_V}
\end{equation}
because $\hat c_\ell|0\rangle=0$ and
$\langle k|\hat c_\ell^\dagger|0\rangle=\delta_{\ell k}$.

Therefore,
\begin{equation}
m_x+i m_y
=
-\sin\theta\,(U_{j,k}+V_{j,k})
\equiv
-\sin\theta\,W_{j,k}.
\label{eq:sm_mperp}
\end{equation}
This defines the combined propagator
\begin{equation}
W_{j,k}\equiv U_{j,k}+V_{j,k}.
\label{eq:sm_Wdef}
\end{equation}

For the orthogonal local encoding
\begin{equation}
\hat R_x^{(k)}(\theta)=e^{-i\theta\hat\sigma_k^x/2},
\label{eq:sm_Rx}
\end{equation}
the encoded state becomes
\begin{equation}
|\Psi_x(\theta)\rangle
=
\cos\frac{\theta}{2}|0\rangle
-i\sin\frac{\theta}{2}\hat c_k^\dagger|0\rangle,
\label{eq:sm_init_Rx}
\end{equation}
and the same calculation gives
\begin{equation}
m_x+i m_y \propto -i\sin\theta\,(U_{j,k}-V_{j,k}),
\label{eq:sm_W_Rx}
\end{equation}
so that the corresponding QFI channel is governed by $|W_{j,k}^{(x)}|^2 = |U_{j,k}-V_{j,k}|^2$.
Thus the two orthogonal encodings probe complementary combinations of the normal
and anomalous propagators.

\subsection{Longitudinal Bloch component}

The parity-even component is
\begin{equation}
  m_z = 1-2\langle\hat n_j(t)\rangle.
\label{eq:sm_mz_def}
\end{equation}
On the vacuum,
\begin{equation}
  \langle0|\hat n_j(t)|0\rangle
  =
  \sum_\ell |V_{j,\ell}|^2
  \equiv S_j,
  \label{eq:sm_Sj}
\end{equation}
which measures pairing-induced occupation of site $j$.
On the one-fermion state,
\begin{equation}
\langle k|\hat n_j(t)|k\rangle
=
S_j + |U_{j,k}|^2 - |V_{j,k}|^2
\equiv S_j+\Delta p_{j,k}.
\label{eq:sm_n_on_k}
\end{equation}
Substituting into Eq.~\eqref{eq:sm_exp_decomp} yields
\begin{equation}
m_z = Z_0+\Delta p_{j,k}\cos\theta,
\label{eq:sm_mz}
\end{equation}
with
\begin{equation}
\Delta p_{j,k}=|U_{j,k}|^2-|V_{j,k}|^2,
\qquad
Z_0=1-2S_j-\Delta p_{j,k}.
\label{eq:sm_aux}
\end{equation}

\subsection{Closed-form QFI and optimal operating point}

Using Eqs.~\eqref{eq:sm_mperp} and \eqref{eq:sm_mz}, the derivatives are
\begin{equation}
\partial_\theta(m_x+i m_y)=-W_{j,k}\cos\theta,
\qquad
\partial_\theta m_z=-\Delta p_{j,k}\sin\theta.
\label{eq:sm_derivatives}
\end{equation}
Hence
\begin{align}
|\partial_\theta\bm m|^2
&=
|W|^2\cos^2\theta+\Delta p^2\sin^2\theta,
\label{eq:sm_speed}\\
\bm m\cdot\partial_\theta\bm m
&=
\sin\theta\Big[\cos\theta(|W|^2-\Delta p^2)-\Delta p\,Z_0\Big],
\label{eq:sm_m_dot_dm}\\
1-|\bm m|^2
&=
1-|W|^2\sin^2\theta-(Z_0+\Delta p\cos\theta)^2.
\label{eq:sm_purityden}
\end{align}
Substituting into Eq.~\eqref{eq:sm_QFI_qubit} gives the exact single-site QFI
quoted in the main text,
\begin{align}\label{eq:sm_QFI_full}
  F_Q^{(j)}=|W|^2\cos^2\theta+\Delta p^2\sin^2\theta+\frac{\sin^2\theta\,[\cos\theta(|W|^2-\Delta p^2)-\Delta p\,Z_0]^2}{1-|W|^2\sin^2\theta-(Z_0+\Delta p\cos\theta)^2}.
\end{align}

At the operating point $\theta_0=0$, the Bloch vector is purely longitudinal,
\begin{equation}
\bm m(\theta_0=0)=(0,0,1-2S_j),
\label{eq:sm_bloch_zero}
\end{equation}
whereas its parameter derivative is purely transverse,
\begin{equation}
\partial_\theta\bm m|_{\theta_0=0}
=
(-\Re W_{j,k},\,-\Im W_{j,k},\,0).
\label{eq:sm_dbloch_zero}
\end{equation}
Therefore
\begin{equation}
\bm m\cdot\partial_\theta\bm m\big|_{\theta_0=0}=0,
\label{eq:sm_orthogonality}
\end{equation}
and Eq.~\eqref{eq:sm_QFI_qubit} collapses to
\begin{equation}
F_Q^{(j)}\big|_{\theta_0=0}
=
|W_{j,k}(t)|^2.
\label{eq:sm_QFI_simple}
\end{equation}
This is the exact small-signal identity used in the main text.

More generally, since the remaining terms in Eq.~\eqref{eq:sm_QFI_full} are
nonnegative,
\begin{equation}
F_Q^{(j)}\ge \cos^2\theta_0\,|W_{j,k}(t)|^2.
\label{eq:sm_QFI_bound}
\end{equation}
The plateau mechanism therefore survives away from the optimal operating point,
with a reduced prefactor.

\section{Majorana zero mode and boundary-axis asymmetry}
\label{sm:majorana}

In the topological phase, the open chain supports a near-zero-energy mode whose
Majorana envelopes are exponentially localized at opposite boundaries. A left
Majorana operator may be written as
\begin{equation}
  \hat\Gamma_L=\sum_j \phi_L(j)\,(\hat c_j+\hat c_j^\dagger),
\label{eq:sm_GammaL}
\end{equation}
and the zero-mode condition $[\hat H,\hat\Gamma_L]=0$ yields the recurrence relation
\begin{equation}
(J+\Delta)\phi_{j+1}+\mu\phi_j+(J-\Delta)\phi_{j-1}=0,
\qquad
\phi_0=0,
\label{eq:sm_recurrence}
\end{equation}
with $\Delta=J\gamma$ and $\mu=-2h$.
Using the ansatz $\phi_j\propto r^{j-1}$ gives the characteristic roots
\begin{equation}
r_\pm=\frac{-\mu\pm\sqrt{\mu^2-4(J^2-\Delta^2)}}{2(J+\Delta)}.
\label{eq:sm_roots}
\end{equation}
The mode is normalizable when $|r_\pm|<1$, which holds in the topological phase
$|\mu|<2J$ for $\Delta\neq0$. The localization length is
\begin{equation}
\xi^{-1}=-\ln|r_>|,
\qquad
r_>\equiv \max(|r_+|,|r_-|).
\label{eq:sm_xi}
\end{equation}

For a finite open chain, the low-energy BdG eigenvector
$(u_0(j),v_0(j))$ can be chosen real. The two Majorana envelopes are then
\begin{equation}
\phi_L(j)=\frac{u_0(j)+v_0(j)}{\sqrt2},
\qquad
\phi_R(j)=\frac{u_0(j)-v_0(j)}{\sqrt2},
\label{eq:sm_phiLR}
\end{equation}
with $\phi_L$ localized at the left boundary and $\phi_R$ at the right
boundary. Their overlap is exponentially small in $L/\xi$.

For the real-coupling chain and $\gamma>0$, the two local encoding channels are
\begin{equation}
W^{(y)}=U+V,
\qquad
W^{(x)}=U-V.
\label{eq:sm_channels}
\end{equation}
Their zero-mode contributions are
\begin{align}\label{eq:sm_W0_Ry}
  W_{j,k}^{(y/x)}(t)\big|_0=2\phi_{L/R}(k)\big[\phi_{L/R}(j)\cos(\varepsilon_0 t)-i\,\phi_{R/L}(j)\sin(\varepsilon_0 t)\big],
\end{align}
At the left boundary, the $\hat R_y$ channel couples to the left Majorana with
$O(1)$ weight, while the $\hat R_x$ channel is controlled by the exponentially small
left-boundary weight of the opposite-edge Majorana. In the thermodynamic or
semi-infinite limit this asymmetry becomes exact; at large finite $L$, the
orthogonal channel contains the exponentially small opposite-edge amplitude $\phi_R(1)\sim e^{-L/\xi}$, and its QFI is therefore exponentially small in $L/\xi$.

If the sign of the real pairing is reversed ($\gamma<0$), the preferred local
axis at a given boundary is interchanged. Throughout the main text and explicit
asymmetry analysis we use the convention $\gamma>0$.

\section{Time-averaged plateau}
\label{sm:plateau}

Separating the zero mode from the gapped bulk, one may write
\begin{equation}
W_{j,k}^{(y)}(t)
=
W_{j,k}^{\rm bulk}(t)
+
W_{j,k}^{(y)}(t)\big|_0.
\label{eq:sm_decomp}
\end{equation}
In the thermodynamic limit, $\varepsilon_0\to0$, so
Eq.~\eqref{eq:sm_W0_Ry} becomes strictly stationary,
\begin{equation}
W_{j,k}^{(y)}\big|_{\rm z.m.}
=
2\phi_L(j)\phi_L(k).
\label{eq:sm_stationary_zm}
\end{equation}
Expanding $|W_{j,k}^{(y)}|^2$ and averaging over time then gives
\begin{equation}
\avg{|W_{j,k}^{(y)}|^2}
=
\avg{|W_{j,k}^{\rm bulk}|^2}
+
4|\phi_L(j)\phi_L(k)|^2,
\label{eq:sm_Wavg}
\end{equation}
because the cross terms oscillate at nonzero frequencies and dephase for generic
open-chain spectra.

At large but finite $L$, the same decomposition is asymptotically exact within
averaging windows satisfying
\begin{equation}
\varepsilon_{\rm gap}^{-1}\ll T_{\rm avg}\ll \varepsilon_0^{-1},
\label{eq:sm_window}
\end{equation}
where $\varepsilon_{\rm gap}$ is the bulk gap and
$\varepsilon_0\sim e^{-L/\xi}$ is the Majorana splitting. The lower bound
ensures dephasing of bulk beats, while the upper bound prevents the near-zero
mode from completing an appreciable oscillation. Thus Eq.~\eqref{eq:sm_Wavg} is
exact in the thermodynamic limit and exponentially accurate at large finite
size.

Using the exact small-signal identity~\eqref{eq:sm_QFI_simple}, the long-time
boundary QFI at the optimal operating point is
\begin{equation}
\avg{F_Q^{(j)}}\big|_{\theta_0=0}
=
\avg{|W_{j,k}^{\rm bulk}|^2}
+
4|\phi_L(j)\phi_L(k)|^2.
\label{eq:sm_plateau_exact}
\end{equation}
This is the plateau formula quoted in the main text.

In the trivial phase, no zero mode exists, so the long-time boundary QFI is
controlled only by the extended bulk modes and vanishes as $O(1/L)$ in the
thermodynamic limit. The plateau therefore requires both a pairing gap and a
topological edge mode.

\subsection{Critical scaling}

For boundary encoding and boundary readout, $j=k=1$, the normalized exponential
Majorana envelope obeys
\begin{equation}
\sum_{j\ge1}\phi_L(j)^2=\frac12.
\label{eq:sm_norm}
\end{equation}
Approximating $\phi_L(j)\propto r_>^{j-1}$ gives
\begin{equation}
\phi_L(1)^2=\frac{1-r_>^2}{2}.
\label{eq:sm_phi1}
\end{equation}
Near the transition at fixed $\gamma\neq0$, one has
$\xi^{-1}=-\ln|r_>| \to 0$, so
\begin{equation}\label{eq:sm_xi_scaling}
  \phi_L(1)^2\sim \xi^{-1},\ \ 
  \xi\sim |h-J|^{-1}\ \ \longrightarrow\ \ \avg{F_Q^{(1)}} \propto |\phi_L(1)|^4 \sim \xi^{-2}\sim |h-J|^2,
\end{equation}
which is the quadratic critical scaling emphasized in the main text. More
generally, at fixed boundary readout and variable encoding site,
\begin{equation}
|\phi_L(1)\phi_L(k)|^2\propto e^{-2(k-1)/\xi},
\label{eq:sm_envelope_scan}
\end{equation}
so the boundary plateau directly resolves the Majorana localization length.

\section{Information velocity}
\label{sm:velocity}

The QFI wavefront propagates with the maximum BdG group velocity,
\begin{equation}
v_{\rm LR}
=
\max_q\left|\frac{d\varepsilon_q}{dq}\right|
=
\max_q \frac{4J|\sin q|\,|h-J(1-\gamma^2)\cos q|}{\varepsilon_q}.
\end{equation}
In the XX limit $(\gamma=0)$ this reduces to $v_{\rm LR}=2J$. In the Ising limit
$(\gamma=1)$ one obtains $v_{\rm LR}=2\min(h,J)$.

\section{Finite-size effects}
\label{sm:finitesize}

At finite $L$, the Majorana splitting scales as $\varepsilon_0\sim
\ee^{-L/\xi}$, so the plateau shows slow oscillations on a timescale $T_{\rm rev}\sim 2\pi/\varepsilon_0$. For small systems ($L\sim 20$) the two edge modes
hybridize strongly and the boundary QFI can periodically collapse to zero. For
larger systems ($L\ge 100$ in our numerics) the oscillations are exponentially
suppressed and the plateau is effectively stable over the time windows used
($tJ\in[150,200]$). The averaging window condition $\varepsilon_{\rm
gap}^{-1}\ll T_{\rm avg}\ll\varepsilon_0^{-1}$ is therefore satisfied.

\section{Locality of the protocol}
\label{sm:locality}

The physical readout protocol is boundary-local. At site $j=1$, the
Jordan--Wigner string is absent, so
\begin{equation}
  \hat c_1+\hat c_1^\dagger=\hat\sigma_1^x,
  \qquad
  -i(\hat c_1-\hat c_1^\dagger)=-\hat\sigma_1^y,
  \qquad
  1-2\hat n_1=-\hat\sigma_1^z.
\label{eq:sm_boundary_local}
\end{equation}
Thus the boundary QFI is obtained from ordinary single-spin Pauli measurements.

Away from the boundary, the fermionic single-site algebra maps to
string-dressed spin operators because of the Jordan--Wigner tail. This is why
the experimentally local protocol considered in the Letter focuses on boundary
readout.

The encoding step is also local in the spin language. On the fully polarized
product state,
\begin{equation}
\hat c_k^\dagger = (-1)^{k-1}\hat\sigma_k^+,
\label{eq:sm_local_encoding}
\end{equation}
so the Jordan--Wigner string contributes only an alternating sign. Since the
QFI is even in the encoded angle, this sign has no effect on the protocol.
Consequently, one may vary the encoding site $k$ while keeping the readout at
$j=1$ and extract the localization length from the exponential dependence of the
boundary plateau, Eq.~\eqref{eq:sm_envelope_scan}, using only local spin
operations throughout.

\end{document}